\documentclass[apj,numberedappendix]{emulateapj}

\usepackage{epsfig}

\def \hinv {{h^{-1}}}



\shorttitle{Cluster Mass Function from Weak Lensing}
\shortauthors{Dahle}
\slugcomment{ApJ in press}

\begin{document}

\title{The Cluster Mass Function from Weak Gravitational Lensing\altaffilmark{1}}

\author{H{\aa}kon Dahle\altaffilmark{2} \\
{\tt hdahle@astro.uio.no}}
\affil{Institute of Theoretical Astrophysics, University of Oslo, \\
P.O. Box 1029, Blindern, N-0315 Oslo, Norway}

\altaffiltext{1}{Based on
observations made with the Nordic Optical Telescope,
operated on the island of La Palma jointly by Denmark, Finland,
Iceland, Norway, and Sweden, in the Spanish Observatorio del Roque de
los Muchachos of the Instituto de Astrof\'\i sica de Canarias}

\altaffiltext{2}{Visiting observer, University of Hawaii 2.24m Telescope at Mauna 
Kea Observatory, Institute for Astronomy, University of Hawaii}

\begin{abstract} {We present the first measurement of the mass function of galaxy clusters based directly on cluster masses derived  
from observations of weak gravitational lensing. To investigate the degree of sample incompleteness resulting from the 
X-ray based selection of the target clusters, we use a sample of 50 clusters with weak lensing 
mass measurements to empirically determine the relation between lensing mass and X-ray luminosity 
and the scatter about this relation. We use a complete, volume-limited sub-sample of 35 X-ray luminous clusters of galaxies at 
$0.15 < z < 0.3$ to constrain the abundance of very massive ($M \gtrsim 10^{15} h^{-1} M_{\sun}$) 
clusters. From this, we constrain $\sigma_8 (\Omega_m /0.3)^{0.37} = 0.67^{+0.04}_{-0.05}$ 
($68\%$ confidence limits), 
agreeing well with constraints from the 3-year WMAP CMB measurements and estimates 
of cluster abundances based on X-ray observations, but somewhat 
lower than constraints from ``cosmic shear'' weak lensing measurements in random fields.}\end{abstract}


\section{Introduction}
\label{sec:intro}

It has long been recognized that the abundance of massive clusters of galaxies 
is a sensitive probe of density fluctuations in the universe. The cluster mass 
function can be used to derive interesting constraints on parameters that govern 
structure growth, in particular the normalization of the primordial 
spectrum of density fluctuations and the average mass density of the universe $\Omega_{m}$ 
(e.g., Evrard 1989; Frenk et al. 1990; Henry \& Arnaud 1991; Lilje 1992; White, Efstathiou, \& Frenk 1993; 
Cen \& Ostriker 1994; Eke, Cole, \& Frenk 1996; Henry 1997; Oukbir \& Blanchard 1997; 
Bahcall \& Fan 1998; Pen 1998; Viana \& Liddle 1999; Henry 2000). 

If the matter-energy density in the universe is dominated by a dark energy component, 
structure growth is hindered when this component becomes dynamically important. 
The resulting effect on the evolution of the cluster mass function makes it an interesting probe of dark energy 
(e.g., Wang \& Steinhardt 1998; Haiman, Mohr, \& Holder 2001; Levine, Schulz, \& White 2002; Munshi, Porciani, \& Wang 2004; Henry 2004; Majumdar \& Mohr 2004). 
Being a sensitive probe of the growth factor, cluster abundances might be combined with probes of the expansion history
such as the magnitude-redshift relation for type Ia supernovae to constrain photon-axion oscillations 
(e.g., Cs{\'a}ki, Kaloper, \& Terning 2005) or even modifications to gravity arising in brane-world cosmologies (Lue, Scoccimarro, \& Starkman 2004).     
Finally, cluster abundances can provide limits on the level of non-Gaussianity in the primordial 
density perturbations (Scoccimarro, Sefusatti, \& Zaldarriaga 2004; Mathis, Diego, \& Silk 2004).   

Many current studies use the X-ray temperature function (XTF) or the X-ray luminosity 
function (XLF) of galaxy clusters as proxies for the mass function (e.g., Pierpaoli et al.\ 2001; Ikebe et al.\ 2002; Reiprich \& B{\"o}hringer 2002; Seljak 2002; Allen et al.\ 2003; Pierpaoli et al.\ 2003; Viana et al.\ 2003; Henry 2004; Stanek et al.\ 2006).
Assuming that the X-ray emitting intra-cluster medium (ICM) is in a state of hydrostatic equilibrium, 
the observed X-ray temperatures $T_{\rm X}$ and luminosities $L_{\rm X}$ can be simply related to cluster masses. 
However, this assumption is not always valid, as some clusters have complex 
temperature structures due to ongoing merging events (e.g., Henry \& Briel 1995; Henriksen \& Markevitch 1996; 
Pratt, B{\"o}hringer, \& Finoguenov 2005). A significant spatial offset between the peaks of the X-ray emission from the ICM and the corresponding concentrations of (collision-less) dark matter is observed in some of these systems
(Clowe, Gonzalez, \& Markevitch 2004; Jee et al.\ 2005). 
Simulations indicate that major mergers will temporarily boost ICM temperatures and luminosities significantly 
above their equilibrium values and thereby significantly alter the observed XTF and XLF 
(e.g., Randall, Sarazin, \& Ricker 2002). This could potentially bias constraints on cosmological parameters, 
if the resulting scatter around the XTF and XLF is not precisely determined for clusters in the high-mass tail 
of the distribution (i.e., if the box sizes of the simulations are too small to adequately sample the most massive 
clusters).   

These problems can be remedied by obtaining independent values for the cluster masses through measurements 
of gravitational lensing, 
which do not rely on assumptions about the dynamical state of the clusters. A set of lensing-derived cluster 
masses can then be used to empirically calibrate the normalization and scatter in the $M-L_{\rm X}$ and $M-T_{\rm X}$   
relations (e.g., Allen et al.\ 2003; Smith et al.\ 2005; Pedersen \& Dahle 2006). 
Still, the complicated gas processes in merging clusters may plausibly lead to a significantly asymmetric 
dispersion about the mean relations which can not be adequately probed by the limited current data sets of 
clusters with both good X-ray and lensing data.  

A different strategy is to derive the cluster mass function directly from lensing masses. 
The main advantage of this approach is that it should in principle avoid the  
problem of relating any observables of the baryonic cluster component to actual cluster mass (in practice, 
however, such issues are still relevant whenever the cluster sample is selected based on baryonic tracers of mass, 
but at a less problematic level). 
This approach requires lensing masses for a large sample of clusters with well-understood selection criteria.  
Samples from cluster surveys that efficiently pick out the extremely rare, most massive clusters in a large volume 
are most suitable, since the high-mass end of the mass function is most sensitive to cosmological parameters. 
Also, given the current level of statistical and systematic uncertainties in lensing measurements, very massive systems 
are needed to produce a useful lensing signal from individual clusters. However, the lensing measurements are 
subject to projection effects that provide an additional source of error, and need to be calibrated to reasonable precision, 
e.g. using N-body simulations (e.g., Clowe, de Lucia, \& King 2004). 

Here, we present the first attempt to constrain the cluster mass function by using weak gravitational lens 
measurements directly. Our results are only dependent on observables of the baryonic component of the clusters 
in the 
sense that the clusters were selected based on their X-ray luminosity. We describe our sample selection in 
\S~\ref{sec:sample}, and in \S~\ref{sec:massfun} we describe in detail how we use our mass measurements to 
derive the mass function. Particular emphasis is put on the treatment of statistical and systematic errors and how 
these affect our results. 

Throughout this paper we will assume a spatially-flat cosmology with $\Omega_m + \Omega_{\Lambda} = 1$. 
Numbers are generally given for the ``concordance model'' case ($\Omega_m$, $\Omega_{\Lambda}$) = ($0.3$, $0.7$). 
The Hubble parameter is generally given by $H_0 = 100 h\, {\rm km}\, {\rm s}^{-1} {\rm Mpc}^{-1}$ (however, in the discussion 
of the $M-L_{\rm X}$ relation we use $H_0 = 50 h_{50}\, {\rm km}\, {\rm s}^{-1} {\rm Mpc}^{-1}$ 
for easy comparison to results in the literature).

\section{Cluster Sample Selection}
\label{sec:sample}
   
Cluster surveys based on photometric and spectroscopic followup of the ROSAT All-Sky Survey (RASS; Tr{\"u}mper 1993) 
using optical telescopes are currently the most efficient strategy for constructing representative 
samples of very massive clusters. 
These surveys cover large volumes, but are not restricted to optically selected objects
such as the clusters in the catalogs of Abell (1958) and Abell, Corwin, \& Olowin (1989), which are seriously 
incomplete in the redshift interval relevant for this work (Briel \& Henry 1993). 

The cluster sample used here was selected from a larger sample of X-ray luminous
clusters of galaxies which were imaged using monolithic $2048^2$ CCD detectors and/or the $8192^2$ UH8K mosaic 
CCD camera. 
We obtained $V$- and $I$-band imaging data for the clusters using the 2.56m Nordic Optical Telescope 
on La Palma, Canary Islands, Spain and the 2.24m University of Hawaii Telescope at 
Mauna Kea Observatory, Hawaii, USA. Exposure times were typically 1.5h in each passband with the $2048^2$
detectors and 3.5h in each passband with the less sensitive UH8K camera. The seeing was in the range 
$0\farcs 6 \leq {\rm FWHM} \leq 1\farcs 1$ in both pass-bands, with a mean seeing of $0\farcs 81$ and $0\farcs 89$
in the $I$-band and $V$-band, respectively. For the weak lensing analysis, background galaxies were selected based on 
signal to noise ratio rather than magnitude, the limits corresponding to the magnitude ranges 
$21 \lesssim m_I \lesssim 24.5$ and $22 \lesssim m_V \lesssim 25.5$ for point sources. The typical number 
density of lensed sources was $\sim 25$ galaxies per square arcminute, and we found a ``figure of merit'' value 
(defined by Kaiser 2000) of $\sum Q^2/d{\Omega} \simeq 1.5 \times 10^5$deg$^{-2}$.
Further details about the observations, data reduction and weak lensing 
analysis are given elsewhere (Dahle et al. 2002; H. Dahle 2006, in preparation). The reduction procedures for 
mosaic CCD data followed the techniques described by Kaiser et al.\ (1999). 
The clusters used here represent a volume-limited sample of X-ray luminous clusters 
taken from the RASS-based, X-ray flux limited ROSAT Brightest Cluster Sample (BCS) of 
Ebeling et al.\ (1998) and its low-flux
extension (eBCS; Ebeling et al.\ 2000). We will hereafter refer to them collectively as the 
(e)BCS sample. We chose a lower cutoff in X-ray luminosity  
of $L_{\rm X, 0.1-2.4 keV} = 6 \times 10^{44}$ erg s$^{-1}$ (limit given for a concordance model 
universe with $h = 0.7$). As discussed in \S~\ref{sec:compl} and illustrated in Figure~\ref{fig:MLrel}, 
this cutoff approximately corresponds to a lower cutoff in mass of $M_{180c} \sim 5 \times 10^{14} \hinv M_{\sun}$. 
Following White (2002), we denote as $M_{180c}$ the mass contained inside the cluster radius $r_{180c}$ 
within which the average density is 180 times the critical density of the universe $\rho_{\rm cr}(z)$ at the 
redshift of the cluster. 

\subsection{Sample Completeness} 
\label{BCScompl}

The estimated overall completeness 
of the BCS sample within its X-ray flux limit is 90\%, and the completeness of the eBCS sample 
is estimated to be 75\% (Ebeling et al.\ 1998; 2000). The selected clusters lie within the redshift range 
$0.15 \leq z \leq 0.303$.
There are 35 (e)BCS clusters that fall within the given limits (and also do not have their 
X-ray signal significantly contaminated by point sources), and we have made weak lensing measurements of all of these (see Table~\ref{tab:BCSsample}). 
Another cluster with a high lensing mass, Cl 1821+643 (Wold et al.\ 2002), is located within the survey volume, 
but it was omitted from the (e)BCS samples because of the presence of a strong AGN point source in the cluster.
B{\"o}hringer et al.\ (2001) find the occurrence of such AGN point source contamination in a large X-ray selected 
cluster sample at similar redshifts to be less than 5\%, so we do not expect to miss more than 1-2 additional 
clusters in our sample from this effect. Furthermore, such cases are accounted for in the BCS and eBCS 
completeness estimates quoted above, and we therefore chose not to include a weak lensing mass measurement for 
Cl 1821+643 in our sample.   

Ebeling et al.\ (1998; 2000) identified the majority of their (e)BCS clusters from the data base produced by running the 
Standard Software Analysis System (SASS) on the RASS data (Voges 1992). The SASS data have a total 
exposure time of 360s with the ROSAT Position Sensitive Proportional Counter (PSPC), this exposure time 
being constant across the sky. Photon event table (PET) files were extracted in a $2\degr \times 2\degr$ region 
centered on each of the SASS-based cluster candidates. In some cases (particularly near the ecliptic poles) the 
PET fields (which contain all the RASS photons collected in the region) are significantly deeper than the SASS data, which 
led to serendipitous discoveries of additional clusters using a Voronoi tessellation and percolation algorithm 
(for full details, see Ebeling et al.\ 1998).  

The BCS and eBCS completeness values quoted above are average values; for both samples there is a small minority 
of clusters that are serendipitous detections, and have much lower completeness values associated with them. 
This is the case for two clusters in our sample (A1576 and ZW5247).    
For each cluster, we list the estimated completenesses, calculated from the completeness associated with the relevant (e)BCS cluster detection method, in  
Table~\ref{tab:BCSsample}.  

\section{Cluster mass function} 
\label{sec:massfun}
   
A number of steps, detailed below, are 
   required to convert weak gravitational lensing measurements for a set of clusters into an observed mass function which can be  
compared to predictions from theory or N-body simulations. Firstly, the total survey volume is calculated for a range of cosmologies, 
and lensing estimates of the cluster mass within a spherical volume are derived. The effects of sample incompleteness
should be taken into account, arising both from incompletenesses in our original cluster catalog (discussed in the preceding section) 
and the mass-dependent incompleteness caused by the observed scatter around the $M_{180c}-L_{\rm X}$ relation, coupled with applying a
sharp lower cutoff in $L_{\rm X}$. Another important correction comes from uncertainties in the cluster mass 
estimates, which will tend to artificially increase the number of clusters in the high-mass tail of the observed mass function. 
    
\subsection{Survey volume}
\label{sec:survey volume}

The (e)BCS survey area consists of the Northern celestial hemisphere ($\delta > 0\degr$), minus a ``zone of avoidance''
at Galactic latitudes $|b| < 20\degr$, where high Galactic extinction may seriously decrease the depth of followup optical extragalactic observations. 
The survey volume is bounded in redshift by $0.15 \leq z \leq 0.303$. For the case of the concordance cosmological model, 
this represents a volume of $8.0 \times 10^8 (h^{-1} {\rm Mpc})^3$. 

\subsection{Cluster mass estimates}
\label{sec:massest}

For our weak lensing analysis, we used the shear estimator of Kaiser (2000). 
This method was tested, along with several other shear estimators, 
by Heymans et al.\ (2006), using simulated weak lensing data.   
The shear estimator of Kaiser (2000) is more mathematically rigorous than the currently 
most widely used shear estimator (Kaiser, Squires, \& Broadhurst 1995), 
but it displays a significant non-linear response to shear, unlike most other shear estimators. 
If we correct our measurements using a second order polynomial in the difference between measured and true 
shear, as found by Heymans et al.\ (2006), 
we find that most cluster masses stay within $+/-15$\% 
of the mass calculated from uncorrected shear values. Furthermore, the change in 
average cluster mass is $<2$\%, i.e., there is very little systematic shift in mass. 
Given the modest size of the correction, and the fact that it would (for a few clusters) require
extrapolations outside the range of shear values over which the shear estimator has 
been tested, we chose not to apply this correction.   

The galaxy shape distortions caused by gravitational lensing provides a measurement of the reduced tangential shear, 
$g_T = \gamma_T / (1 - \kappa )$, where $\gamma_T$ is the tangential component of the shear and $\kappa$ is the convergence.   
The average value of $g_T$ is measured in a set of non-overlapping annuli with mean radius $r$, centered on the mass center of the cluster.
We fit a NFW-type profile (Navarro, Frenk, \& White 1997) to the observed shear profile $g_T(r)$
of each cluster. In this model, the density is $\rho (r) \propto \left[ (r/r_s)(1+r/r_s)^2 \right]^{-1}$, and it is 
specified by a scale radius $r_s$ and a concentration parameter $c_{\rm vir} = r_{\rm vir}/r_s$, where $r_{\rm vir}$
is the virial radius of the cluster. N-body simulations of clusters in a $\Lambda$CDM universe predict the dependency 
of $c_{\rm vir}$ on cluster mass and redshift (Bullock et al.\ 2001; Dolag et al.\ 2004). 
We have previously shown that the NFW model provides a good fit to the 
average shear profile of a sample of 6 galaxy clusters at $z\sim 0.3$ (Dahle, Hannestad, \& Sommer-Larsen 2003), 
with a concentration parameter consistent with predictions. Studies based on X-ray observations reach similar  
conclusions (Pointecouteau, Arnaud, \& Pratt 2005; Vikhlinin et al.\ 2006; Voigt \& Fabian 2006). 
As shown in Table~\ref{tab:BCSsample}, the median cluster mass of our sample is $M_{180c} \simeq 8\times 10^{14} M_{\sun}$, 
for which Bullock et al.\ (2001) predict a median halo concentration of $c_{\rm vir} = r_{\rm vir} / r_s = 1.14 r_{180c} / r_s = 5.8 / (1+z)$ 
in a concordance model universe. Given the modest range in masses for the clusters studied here, the weak 
mass-dependence ($c_{\rm vir} \propto M^{-0.13}$) predicted by these authors is negligible, and a fixed value for 
$c_{\rm vir}$ is adopted at a given redshift. For derivations of the lensing properties of the NFW model, 
we refer the reader to the papers of Bartelmann (1996) and Wright \& Brainerd (2000).  

The NFW model fit gives estimates of $r_{180c}$ and $M_{180c}$.  
The shear measurements used for the fit were made at cluster-centric radii $50\arcsec < r < 180\arcsec$ 
for clusters which 
were observed with $2048^2$ CCD cameras and $150\arcsec < r < 550\arcsec$ for the clusters which were 
observed with the UH8K mosaic CCD camera. 
For our clusters, we derive $r_{180c}$ values typically corresponding to an angular extent in the range $450\arcsec < r_{180c} < 750\arcsec$. 
Hence, extrapolations were in many cases required to derive the masses within $r_{180c}$. 
As we assume a fixed value for $c_{\rm vir}$ at a given redshift, any intrinsic scatter in this parameter will introduce an
extra uncertainty in the cluster mass estimates. The level of dispersion
predicted by Bullock et al.\ (2001) and Dolag et al.\ (2004) 
(a $1 \sigma$ scatter of $\Delta (\log c_{\rm vir}) \sim 0.18$) gives a contribution to 
the overall error budget for the cluster masses which depends on the ratio of the maximum radius of the shear 
measurements $r_{\rm fit}$, to $r_{180c}$, which is listed in Table~\ref{tab:BCSsample}. 
For the clusters observed with $2048^2$ CCD cameras ($r_{\rm fit}/r_{180c} \simeq 0.3$), the predicted scatter in $c_{\rm vir}$ results in an 
additional mass uncertainty of 18\%, which was assumed to be unrelated to other measurement uncertainties.
Similarly, for the clusters observed with the UH8K camera ($r_{\rm fit}/r_{180c} \simeq 1.0$), 
an uncertainty of 10\% was estimated and added in quadrature to other contributions to the mass measurement 
uncertainty (see below).

\begin{deluxetable*}{lcrrccc}
\tablecaption{Weak lensing masses and X-ray luminosities
  \label{tab:BCSsample}}
\tablehead{
\colhead{Cluster} &
\colhead{Redshift\rlap{\tablenotemark{a}}} &
\colhead{L$_{\rm X, 0.1-2.4 keV}$\rlap{\tablenotemark{b}}} &
\colhead{$M_{180c}$} &
\colhead{$\epsilon (M_{180c})$} &
\colhead{$r_{\rm fit}/r_{180c}$} &
\colhead{(e)BCS} 
\\
\colhead{name} & \colhead{$z$} & \colhead{($h_{50}^{-2}$ $10^{44} {\rm erg~s}^{-1}$)} & \colhead{($10^{14} h^{-1} M_{\sun}$)} & \colhead{($10^{14} h^{-1} M_{\sun}$)} & & \colhead{completeness}
}
\startdata
           \objectname{A2204} \dotfill &          0.15 &     24.6 &          8.36 &          5.19 & 0.22    & 0.90 \\
         \objectname{RX J1720.1+2638} \dotfill &  0.16 &     18.8 &          4.66 &          3.37 & 0.29    & 0.90 \\
            \objectname{A586} \dotfill &          0.17 &     13.0 &         26.80 &          8.89 & 0.17    & 0.90 \\
          \objectname{A1914} \dotfill &           0.17 &     21.6 &          7.20 &          4.41 & 0.26    & 0.90 \\
            \objectname{A665} \dotfill &          0.18 &     19.3 &          8.27 &          4.38 & 0.26    & 0.90 \\
            \objectname{A115} \dotfill &          0.20 &      6.9 &          6.01 &          4.14 & 0.31    & 0.90 \\
            \objectname{A520} \dotfill &          0.20 &     17.4 &         13.03 &          4.48 & 0.75    & 0.90 \\
           \objectname{A963}  \dotfill &          0.21 &     12.6 &          6.65 &          5.81 & 0.95    & 0.90 \\
           \objectname{A1423} \dotfill &          0.21 &     12.1 &         11.80 &          5.94 & 0.26    & 0.90 \\
            \objectname{A773} \dotfill &          0.22 &     16.0 &         12.07 &          5.38 & 0.27    & 0.90 \\
           \objectname{A2261} \dotfill &          0.22 &     22.2 &          6.18 &          3.58 & 0.34    & 0.90 \\
            \objectname{A267} \dotfill &          0.23 &     16.9 &         11.80 &          4.36 & 0.86    & 0.90 \\
           \objectname{A1682} \dotfill &          0.23 &     13.8 &          4.38 &          2.92 & 0.39    & 0.90 \\
           \objectname{A1763} \dotfill &          0.23 &     18.4 &          8.34 &          4.02 & 0.31    & 0.90 \\
           \objectname{A2111} \dotfill &          0.23 &     13.4 &          4.88 &          2.85 & 0.38    & 0.90 \\
           \objectname{A2219} \dotfill &          0.23 &     25.1 &          6.50 &          4.44 & 0.34    & 0.90 \\
           \objectname{A2390} \dotfill &          0.23 &     26.4 &         11.99 &          5.28 & 0.28    & 0.90 \\
          \objectname{Zw 5247} \dotfill &         0.23 &     12.4 &          2.53 &          2.09 & 0.47    & 0.15 \\
         \objectname{RX J2129.6+0005} \dotfill &  0.23 &     22.9 &          5.97 &          4.09 & 0.36    & 0.90 \\
         \objectname{RX J0439.0+0715} \dotfill &  0.24 &     16.5 &         10.00 &          5.08 & 0.31    & 0.78 \\
          \objectname{Zw 2089} \dotfill &         0.24 &     13.3 &          3.64 &          3.04 & 0.43    & 0.90 \\
          \objectname{A1835} \dotfill &           0.25 &     48.2 &          8.39 &          4.43 & 0.34    & 0.90 \\
             \objectname{A68} \dotfill &          0.26 &     18.7 &         17.08 &          9.01 & 0.27    & 0.90 \\
          \objectname{MS1455+22} \dotfill &       0.26 &     16.6 &         10.16 &          4.71 & 0.33    & 0.90 \\
         \objectname{Zw 5768} \dotfill &          0.27 &     14.7 &          6.04 &          4.65 & 0.40    & 0.78 \\
            \objectname{A697} \dotfill &          0.28 &     20.8 &         20.40 &          7.34 & 0.28    & 0.90 \\
           \objectname{A1758N} \dotfill &         0.28 &     14.9 &         21.02 &          7.36 & 0.27    & 0.90 \\
           \objectname{A2631} \dotfill &          0.28 &     16.8 &          4.88 &          3.33 & 0.44    & 0.78 \\
            \objectname{A611} \dotfill &          0.29 &     17.4 &          5.21 &          3.47 & 0.29    & 0.78 \\
         \objectname{RX J0437.1+0043} \dotfill &  0.29 &     15.7 &          3.74 &          2.65 & 0.49    & 0.78 \\
          \objectname{Zw 3146} \dotfill &         0.29 &     34.0 &          8.94 &          4.23 & 0.38    & 0.90 \\
          \objectname{Zw 7215} \dotfill &         0.29 &     14.4 &          8.38 &          4.67 & 0.38    & 0.78 \\
            \objectname{A781} \dotfill &          0.30 &     22.2 &         10.00 &          4.55 & 0.37    & 0.78 \\
           \objectname{A1576} \dotfill &          0.30 &     14.1 &         14.49 &          4.66 & 1.00    & 0.13 \\
           \objectname{A2552} \dotfill &          0.30 &     19.6 &          3.70 &          2.91 & 0.52    & 0.78 \\
\enddata
\tablecomments{The luminosity and lensing mass values are given for a flat cosmology with $\Omega_m = 0.3$, $\Omega_{\Lambda} = 0.7$. The tabulated mass uncertainties include an assumed 26\%  from projection effects, added in quadrature to the lensing mass estimate. The tabulated errors are calculated by symmetrising the error bars.} 
\tablenotetext{a}{--- See Ebeling et al.\ (1998; 2000) for references to redshift measurements.}
\tablenotetext{b}{--- The X-ray luminosity values are taken from Ebeling et al.\ (1998; 2000).
}
\end{deluxetable*}

Our photometric data in two pass-bands did not allow reliable discrimination between 
cluster galaxies and lensed background galaxies (except for the small fraction of galaxies that had 
observed $V-I$ colors redder than early-type cluster galaxies at the cluster redshift).
The faint galaxy catalogs which were used to measure the gravitational shear would thus be significantly 
contaminated by cluster galaxies that will dilute the lensing signal by an amount 
given by their local sky density.   
To correct for this contamination, a radially dependent correction factor was applied to the shear.  
The magnitude of the correction was estimated based on the radial dependence of the average 
faint galaxy density in two different sets of clusters observed with the UH8K camera, one at $z\sim 0.3$ and 
the other at $z\sim 0.2$, assuming that the contamination is negligible at the edge of the UH8K fields. 
This procedure led to corrections to the cluster masses at the 20-30\% level (see Pedersen \& Dahle 2006 
for details). Cluster-to-cluster variations in their richness in faint dwarf galaxies will lead to an additional 
scatter when a fixed level of contamination is assumed for all clusters. Based on our data, we estimate a typical 
scatter in cluster dwarf richness of 50\%, leading to an additional uncertainty in the cluster mass measurements 
of 12\% for data obtained with with the UH8K camera (which include lensing measurements at large radii 
where the contamination is less severe) and 20\% for the clusters observed with the $2048^2$ CCD cameras. 
These were added in quadrature to the mass measurement uncertainties.   

The redshift distribution of the background galaxies 
needs to be known to convert the observed lensing signal into a cluster mass estimate. 
The background galaxy redshifts were estimated from spectroscopic and photometric redshifts in 
the Hubble Deep Field (for details, see Dahle et al.\ 2002). The average value 
of the ratio between the lens-source and observer-source angular diameter distances, $\beta \equiv D_{ls}$/$D_s$, 
was approximated by a relation on the form $\langle \beta \rangle = A z_{\rm cl}^2 + B z_{\rm cl} + C$, 
where $z_{\rm cl}$ is the cluster redshift. The constants were calculated for a range of spatially-flat cosmologies; 
for the concordance model, we obtained $A=1.37$, $B=-2.00$, and $C=1.01$. 

Gravitational lensing effects measure the projected mass distribution in the cluster, including mass outside $r_{180c}$.
This may introduce a bias and a dispersion in the derived lensing estimates of $M_{180c}$.  
Based on N-body simulations of clusters in a $\Lambda$CDM model, Metzler, White \& Loken (2001) considered a lensing mass estimator based on the 
projected mass within $r_{200b}$ and $r_{500b}$ (the radii within which the mean density contrast is $\bar{\delta} = 200$ and $\bar{\delta} = 500$, respectively). 
For their estimator, they found that lensing masses would on average be biased by a factor $\langle M_{\rm lens}/M_{\rm true} \rangle = 1.33$, 
with a mass dispersion of 0.26 around the mean. However, these authors remarked that this positive bias could largely be removed by 
assuming a realistic model for the radial mass distribution out to large radii.   
A more recent study (Clowe, De Lucia, \& King 2004) indicated that there is on average no significant bias
in the lensing-derived masses, provided that these are determined by fitting the observed tangential shear to an NFW model, 
using shear measurements all the way out to $R_{180c}$. Still, a significant scatter will be added to the derived  
lensing masses, and we added such a scatter in quadrature to the measurement uncertainties of our lensing mass estimates. 
We performed our calculations using three different values of the scatter (0.26, 0.13, and 0), but found only a weak 
effect on the end result, as the shear measurement uncertainties dominate the error budget for most of the clusters.
All subsequent results in this paper are based on adopting the highest value (26\%) of this scatter.  

\subsection{Mass-X ray luminosity relation}
\label{sec:compl}

The observed scatter around the $M_{180c}-L_{\rm X}$ relation causes a mass-dependent sample incompleteness which is most severe at the
lower end of the range of measured cluster masses. 
Our sample is defined by a fixed cutoff in X-ray luminosity which corresponds to $L_{\rm X, 0.1-2.4 keV} = 1.2 \times 10^{45} h_{50}^{-2}$ erg s$^{-1}$ 
for the concordance model. Any intrinsic scatter around the mean $M_{180c}$ - $L_{\rm X}$ relation and uncertainties in the X-ray luminosity 
measurements tabulated by Ebeling et al.\ (1998; 2000) will contribute to soften the corresponding cutoff in mass.    
To assess the completeness of our sample as a function of $M_{180c}$, we made an empirical determination of the $M_{180c}$ - $L_{\rm X}$
relation and its scatter.  
This was done using a data set of 50 clusters (Dahle et al.\ 2002; H. Dahle 2006, in preparation) which, in addition to the 35 clusters 
used here for the determination of the mass function, include clusters of similarly high X-ray luminosities, selected from four cluster samples 
(Ebeling et al.\ 1996, 1998, 2000; B{\"o}hringer et al.\ 2000), all based on RASS data.  
We adopted the RASS-based X-ray luminosities listed by these authors (with preference for the (e)BCS values whenever a cluster 
is included in more than one sample), after scaling them to our chosen cosmological model. 

For many of these clusters, more precise $L_{\rm X}$ values exist in the literature, e.g., from pointed observations with ROSAT, XMM-Newton 
or {\it Chandra}. However, our current aim is not to determine the $M_{180c}-L_{\rm X}$ with the highest possible accuracy, but 
rather to quantify the incompleteness effects caused by selecting clusters based on their X-ray luminosities in the RASS cluster surveys. 
$L_{\rm X}$ only serves as a proxy for mass for our cluster sample selection, and when determining (below) the scatter about the mean $M_{180c}-L_{\rm X}$-relation 
we do not distinguish the contribution to the scatter caused by uncertainties in the $L_{\rm X}$-measurements from the intrinsic scatter 
about the relation, as they will both act to soften the low-mass cutoff in a similar way.  

\begin{figure*}
\centering\epsfig{file=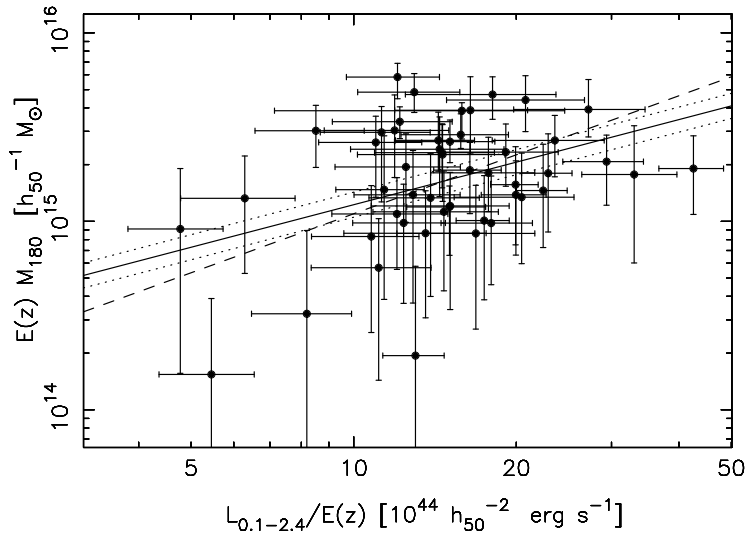,width=7cm,angle=-90}
\caption[M-L_X relation.]
{Weak lensing mass estimates of 50 clusters plotted against their X-ray luminosities. The dashed line indicates the best fit to the mass-luminosity relation, 
with an arbitrary slope $\alpha$. Also shown is the best-fit relation when assuming the theoretically expected value $\alpha = 0.75$ (solid line). 
The dotted lines represent the $1\sigma$ uncertainty about this relation.} 
\label{fig:MLrel}
\end{figure*}

The lensing masses and X-ray luminosities of the clusters are plotted in Figure~\ref{fig:MLrel}. The large apparent spread in this figure, compared to 
similar diagrams from X-ray data, is mostly a result of the narrow range in $L_{\rm X}$, and hence, in cluster mass.

We assumed a $M_{180c}-L_{\rm X}$ 
relation on the form 

\begin{equation} 
E(z) M_{180c} = M_0  \left[\frac{L_{\rm X}}{E(z)}\right]^{\alpha}, 
\label{eq:theoryML}
\end{equation} 

\noindent
where $E(z) = (1 + z) \sqrt{(1 + z\Omega_m + \Omega_{\Lambda} / (1+z)^2 - \Omega_{\Lambda})}$ is an evolution parameter
that describes the redshift-dependence predicted from simulations (e.g., Bryan \& Norman 1998; Mathiesen \& Evrard 2001). 
Using the BCES($X_2|X_1$) estimator of Akritas \& Bershady (1996), we fitted the data in Figure~\ref{fig:MLrel} to a model on the form

\begin{eqnarray} 
\log \left[\frac{E(z) M_{180c}}{h_{50}^{-1} M_{\sun}}\right] = \nonumber \\ \alpha \log \left[\frac{L_{\rm X}}{E(z) 10^{44} h_{50}^{-2} {\rm erg\, s}^{-1} }\right] +  \log \left(\frac{M_0}{h^{-1}_{50} M_{\sun}}\right). 
\label{eq:theoryML2}
\end{eqnarray} 

This estimator takes into account the errors in both axes, and allows for possible intrinsic scatter 
in the $M_{180c}-L_{\rm X}$ relation. 
Our best-fit values are $\alpha = 1.04 \pm 0.46$ and $\log \left(M_0/h^{-1}_{50} M_{\sun}\right) = 14.0 \pm 0.6$.
As can be expected from Figure~\ref{fig:MLrel}, the slope $\alpha$ is rather poorly constrained, since the clusters 
in our sample only span modest ranges in $M_{180c}$ and $L_{\rm X}$. Figure~\ref{fig:MLrel} also seems 
to indicate a flattening of the $M_{180c}-L_{\rm X}$ slope at very high X-ray luminosities 
($L_{\rm X, 0.1-2.4 keV} \gtrsim 2 \times 10^{45} h_{50}^{-2}$ erg s$^{-1}$), similar to the change   
of the slope of the $M-T_{\rm X}$ relation at high temperatures noted by Cypriano et al.\ (2004). 
In both cases, this is most likely caused by a scattering of out-of-equilibrium clusters into the extreme 
tails of the XTF and XLF, as discussed in \S~\ref{sec:intro}.  
  
We also tried fixing the slope parameter at the value $\alpha = 0.75$ predicted from a self-similar collapse 
model for galaxy clusters (Kaiser 1986). 
Our fit to the relation in Eq.~\ref{eq:theoryML} then gave a normalization 
$\log (M_0/h_{50}^{-1} M_{\sun}) = 14.34 \pm 0.07$. 
This is consistent with the relation similarly derived from a sample of 17 clusters by Allen et al.\ (2003). 
These authors used a combination of X-ray based mass measurements derived from {\it Chandra} data for a set of 10 clusters they had classified as 
relaxed systems, and weak lensing-based mass measurements for a partially overlapping sample of 10 clusters, including some unrelaxed systems. 
Since $L_{\rm X}$ is our only cluster selection criterion, we have not here attempted to classify clusters according to their dynamical state.  

When accounting for the additional scatter about this relation caused by the mass measurement uncertainties, a scatter of $\sigma_M = 0.44$ is found. 
By comparison, the data presented by Reiprich \& B\"ohringer (2002) indicate an intrinsic scatter of $\sigma_M = 0.39$ about the $M-L_{\rm X}$ relation, using X-ray based 
mass estimates under the assumptions of isothermality and hydrostatic equilibrium of the intra-cluster gas. More recently, 
Smith et al.\ (2005) found an intrinsic scatter $\sigma_M = 0.41$ about their $M-L_{\rm X}$ relation, based on gravitational lensing 
measurements of the masses of 10 clusters within a cluster-centric radius of $250 h^{-1}$\, kpc. The slightly larger scatter we measure could be a result of 
 the larger uncertainties of $L_{\rm X}$ values determined from RASS data, compared   to deeper pointed X-ray observations of clusters.

For a given cluster of true mass $M_{180c}$, the probability of being included in our sample is

\begin{equation} 
P(M_{180c}) = \int_{- \infty}^{M_{180c}} P_{\delta M_L} (M - M(L_{\rm min})) dM, 
\label{eq:probincl}
\end{equation} 

\noindent
where $P_{\delta M_L}$ is the probability distribution of $M_{180c}$ for a given X-ray luminosity and $M(L_{\rm min})$ 
is the mass corresponding (via eq.~\ref{eq:theoryML}) to the cutoff in X-ray luminosity $L_{\rm min} = 1.2 \times 10^{45} h_{50}^{-2}$ erg s$^{-1}$. 
Here, $P_{\delta M_L}$ is assumed to be a log-normal function (analogous to the scatter about the X-ray luminosity-temperature relation; see Novicki, Sornig, \& Henry 2002), 
for which we derive a standard deviation in $\Delta \log M$ of $\sigma_{\log M} = 0.178$, based on the data plotted in Figure~\ref{fig:MLrel}.  
We note that the exact form of $P_{\delta M_L}$ should be determined empirically in the future, using even larger data sets of X-ray luminous 
clusters with more accurately determined lensing masses.  

To make our subsequent analysis self-consistent, the appropriate value of $M(L_{\rm min})$ was determined for the range of spatially-flat cosmologies probed in this study; its value at the mean cluster 
redshift ($z=0.23$) for the concordance model 
is $M(L_{\rm min}) = 5.2 \times 10^{14} \hinv M_{\sun}$. 

\subsection{Measurement uncertainties} 
\label{sec:uncert}

A significant uncertainty is associated with any realistic cluster mass estimator. In the case of gravitational lensing-based mass measurements, uncertainties arise both from the measurement uncertainties of the observable itself and from the intrinsic scatter in the mass-observable relation caused by projection effects. As discussed in \S~\ref{sec:massest}, the former is dominant for 
most of the clusters in our data sample, but there is also a non-negligible contribution from the latter. Given these uncertainties, extra care must be taken to remove possible biases 
produced by the exponential falloff of the mass function at high masses. If the uncertainties in $M_{180c}$ were the same for all the clusters, 
they could simply be incorporated by convolving the theoretical mass function with a probability distribution $P(\delta M_{180c})$, with a width 
given by the uncertainties (e.g., Evrard 1989; Metzler et al.\ 2001). However, as shown in Table~\ref{tab:BCSsample}, the uncertainties vary significantly  
from cluster to cluster. Hence, the theoretical mass function is convolved with the uncertainty of each cluster, and the average of the resulting ensemble of mass functions is calculated and used in our further analysis.

Given a theoretical mass function $n_{\rm th}(M_{180c})$, and given the total measurement uncertainties in the weak lensing estimates of $M_{180c}$, 
the observed mass function can be predicted as:   
    
\begin{eqnarray} 
n_{\rm obs} (M_{180c})=     \nonumber \\ 
  \left< \int d \delta M_{180c} P(\delta M_{180c}) n_{\rm th}(M_{180c} + \delta M_{180c}) P(M_{180c}) \right> , \nonumber
\label{eq:obsmassfun}
\end{eqnarray} 

\noindent
where $P(\delta M_{180c})$ is a log-normal probability distribution of width given by the measurement uncertainty in $M_{180c}$, and the average is taken over 
the ensemble of observed uncertainties $\sigma_{M180c}$.  

Finally, we note that $M_{180c}$, $\sigma_{M180c}$, and $M(L_{\rm min})$ are all cosmology-dependent quantities, and to make our analysis  
self-consistent, these need to be re-calculated for each cosmology.   

\begin{figure*}
\centering\epsfig{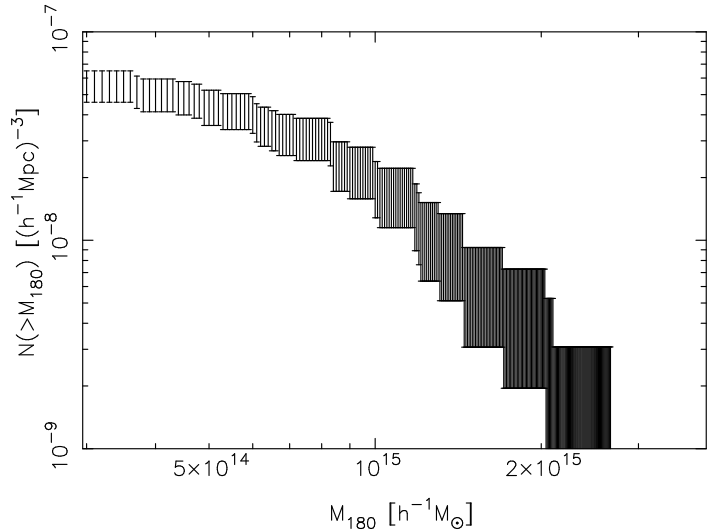}
\caption[Observed mass function.]
{Observed cumulative mass function. The apparent flattening of the slope at $\lesssim 8 \times 10^{14} \hinv M_{\sun}$ is caused 
by severe incompleteness of the sample at lower masses. This plot is shown for illustrational purposes only; the actual procedure  
used for fitting the data to theoretical mass functions is described in \S~\ref{sec:thfit}.} 
\label{fig:Massfun}
\end{figure*}

\subsection{Fit to predicted mass functions} 
\label{sec:thfit}

A number of studies have predicted the mass function of dark matter halos using an analytic approach or via numerical simulations. 
The formalism of Press \& Schechter (1974) uses linear theory for the evolution of self-gravitating mass density fluctuations and assumes 
a Gaussian distribution of inhomogeneities to calculate the fraction of the volume which has reached a given over-density for a given 
smoothing scale, assuming that all objects at a spherical over-density above a threshold value $\delta_c = 1.69$ undergo gravitational collapse.
The mass function derived via Press-Schechter theory provides overall a reasonable match to mass functions derived from N-body simulations of CDM particles, 
over a wide range of masses (e.g., White et al.\ 1993). However, the mass functions derived from simulations still show significant departures from 
the Press-Schechter form, yielding a larger number of clusters at the high-mass end (e.g., Jenkins et al.\ 2001). Sheth, Mo \& Tormen (2001) showed that a 
modification of Press-Schechter theory to allow for ellipsoidal, rather than spherical, collapse gave a significantly improved fit to 
the simulations, similar to the form suggested by Sheth \& Tormen (1999). Recently, Warren et al.\ (2006) used a new, larger set of simulations and found  
that the mass functions introduced by Jenkins et al.\ (2001) and Sheth \& Tormen (1999) deviate by $> 30\%$ from their simulations at the 
highest masses. They therefore provide a new formula for the mass function shape. 

Based on these results, we chose to fit our cluster data to two mass function forms; one which is consistent with the most recent simulations
(Warren et al.\ 2006), and one which is more directly based on analytical work (Sheth \& Tormen 1999).
Jenkins et al.\ (2001) and White (2002) showed that the shape of the mass function is universal across the range of currently interesting cosmologies, 
given that the halo mass ($M_{180b}$) is always defined within a volume of average over-density $\bar{\delta} = 180$ with respect to the mean (rather than critical) 
matter density of the universe. In the case of the simulations, Warren et al.\ (2006) remark that the commonly used ``friends-of-friends'' halo definition 
method tends to select halos of a different average over-density, which depends on the number of particles in the simulated halo. For the most massive halos, the halo 
mass is $M_{280b}$, i.e., corresponding to an average halo over-density $\Delta = 280$. We convert the mass function predictions of Warren et al.\ (2006) 
and Sheth \& Tormen (1999) to $M_{180c}$ (from $M_{280b}$ and $M_{180b}$, respectively), using the fitting function in Appendix C of Hu \& Kravtsov (2003), based on the procedure outlined by White (2001).  
We assume a structure parameter fixed at $\Gamma = 0.21$. 

For easy comparison between theory and observations, we binned the theoretical mass function, such that the predicted number of clusters in the 
i'th bin would be  

\begin{equation} 
N_{i} = V_{\rm BCS} \int_{m_{i,1}}^{m_{i,2}} \frac{dn}{dM} dM, 
\label{eq:binning}
\end{equation} 

\noindent  
where $V_{\rm BCS} (\Omega_m, \Omega_{\Lambda} = 1 - \Omega_m)$ is the BCS survey volume defined in \S~\ref{sec:survey volume}. 
For the theoretical mass functions, we calculated the predicted number of observed clusters in four mass bins (of widths 5, 5, 5, and $10 \times 10^{14} \hinv M_{\sun}$, respectively), ranging from  
$2.3 \times 10^{14} \hinv M_{\sun} < M_{180c} < 7.3 \times 10^{14} \hinv M_{\sun}$ to $17.3 \times 10^{14} \hinv M_{\sun} < M_{180c} < 27.3 \times 10^{14} \hinv M_{\sun}$. 
In the end, we only used the three uppermost mass bins, since the uncertainties of the lowest mass 
bin will be very large, and completely dominated by uncertainties in the incompleteness correction, 
given the small degree of sample completeness in this mass range. In the lowest bin, the 
incompleteness correction will be highly sensitive to the scatter around the mean 
$M_{180c}-L_{\rm X}$ relation. As shown in \S~\ref{sec:compl}, the sample selection cutoff in 
$L_{\rm X}$ corresponds to a lower cutoff in mass of $M_{180c} \sim 5 \times 10^{14} \hinv M_{\sun}$.  
To account for Poisson 
shot noise, $\sqrt{N}$ errors are applied to the counts of clusters in each mass bin. Additional 
uncertainties arise from the cluster mass measurement uncertainties, resulting in a non-negligible 
probability that a cluster is counted in the wrong bin. These errors are added in quadrature to the 
Poisson errors. In addition, $\sqrt{N}$ errors are applied to the theoretically predicted cluster 
counts in each bin, to account for cosmic variance.   
Fitting to the theoretical mass function of Sheth \& Tormen (1999), we find 
$\sigma_8 (\Omega_m /0.3)^{0.37} = 0.67^{+0.04}_{-0.05}$ ($68\%$ confidence limits), 
while a similar fit to the mass function of Warren et al.\ (2006) gives 
$\sigma_8 (\Omega_m /0.3)^{0.37} = 0.66\pm 0.04$. 

There is a tendency for a preference for higher $\sigma_8$ values in the higher mass bins; the uppermost 
mass bin alone gives a best fit $\sigma_8 = 0.72^{+0.05}_{-0.10}$ for $\Omega_m = 0.3$, using the mass function of Sheth \& Tormen (1999). This may hint towards an
overestimate of the sample completeness in the lower mass bins. Stanek et al.\ (2006) discusses the effect of Malmquist bias 
on X-ray flux-limited cluster samples in the presence of large intrinsic scatter around the $M - L_{\rm X}$ relation 
and suggest that the true normalization of this relation is a factor of 2 dimmer than estimated from the RASS-based 
HIFLUGCS sample of Reiprich \& B{\"o}hringer (2002). Applying a shift of this magnitude to the $M_{180c} - L_{\rm X}$    
relation derived in \S~\ref{sec:compl} (while keeping the estimated level of scatter, which is consistent with the  
scatter estimated by Stanek et al.\ 2006) would increase the value of $M(L_{\rm min})$ (to $8.7 \times 10^{14} \hinv M_{\sun}$, in the case of the concordance model) and increase the incompleteness correction factors in each mass bin, yielding a best fit 
$\sigma_8 (\Omega_m = 0.3) = 0.74^{+0.04}_{-0.06}$, based on the three uppermost mass bins. However, 
since the sample of 50 clusters used to derive the $M_{180c} - L_{\rm X}$ relation in 
\S~\ref{sec:compl} can be more accurately described as a volume-limited rather than flux-limited 
sample, this Malmquist bias correction should not be applicable to our sample. 

\begin{figure*}
\centering\epsfig{file=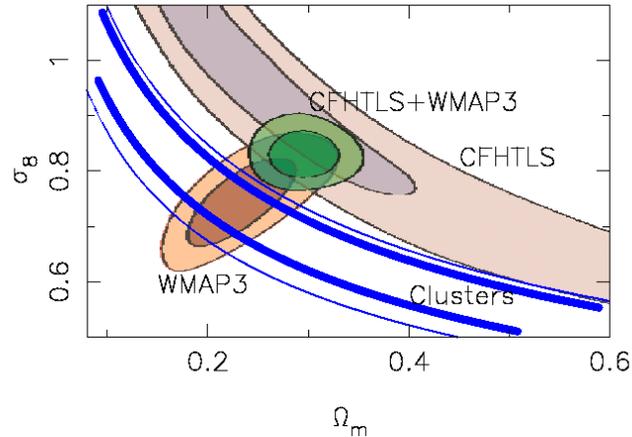,width=8cm,angle=-90}
\caption[Constraints.]
{Joint constraints on $\sigma_8$ and $\Omega_m$. The blue curves show 68\% (thick lines) and 95\% (thin lines) constraints on these two parameters from the galaxy cluster data presented in this paper.   Also shown are constraints on the same parameters from the CFHTLS weak lensing survey, from the 3-year WMAP CMB data, and joint constraints from these two data sets, derived by Spergel et al.\ (2006).} 
\label{fig:bananaSpergel}
\end{figure*}

It should be noted that the magnitude of the uncertainties in the gravitational lensing mass measurements 
of the clusters, including the three additional sources of mass scatter discussed in \S~\ref{sec:massest}, 
has a significant effect on the derived cosmological parameters (see e.g., Lima \& Hu 2005 for a discussion 
of a similar effect on dark energy constraints from high-redshift clusters). The effect of the scatter is to 
populate the high-mass tail of the cluster mass function with some of the 
(far more numerous) lower-mass clusters which have been scattered upwards in mass. As an example, 
if we artificially reduce the scatter in mass in our data by 30\% (as would be the effect of ignoring 
the effect of mass scatter from variations in the NFW concentration parameter, variations in cluster galaxy 
contamination and projection effects), the best fit value of $\sigma_8$ for the concordance model would 
increase from 
$\sigma_8 = 0.67^{+0.04}_{-0.05}$ to $\sigma_8 = 0.70^{+0.04}_{-0.05}$. Hence, a decrease in the mass 
uncertainties works  
in the opposite direction as an decrease in the mass values, so that these will partially cancel each other 
out if the cluster masses and mass uncertainties are multiplied by the same factor. This makes our $\sigma_8$ 
result fairly robust to e.g.\ a modest multiplicative error in the shear estimator.   

\section{Discussion}
\label{sec:disc}

The results presented in this paper demonstrate that weak gravitational lensing based 
measurements of cluster abundances provide competitive constraints 
in the $\sigma_8$-$\Omega_m$ plane. In a recent work (Pedersen \& Dahle 2006), $\sigma_8$ was estimated 
by measuring the normalization of the mass-X ray temperature relation of galaxy clusters.
If the intrinsic scatter about the mass-temperature relation 
is low (10\% or less), as this normalization implies a higher value of $\sigma_8 = 0.88 \pm 0.09$ for 
$\Omega_m = 0.3$. While most theoretical studies predict a rather tight mass-temperature 
relation, Pedersen \& Dahle (2006) showed that the scatter about the relation is most likely larger than 10\% , producing a
$\sigma_8$ estimate which is systematically biased towards high values. Hence, as remarked in their discussion, 
the $\sigma_8$ estimate of Pedersen \& Dahle (2006) should most accurately be regarded as an upper limit, and the true value is likely to be lower.  

Being based directly on weak gravitational lensing mass estimates, the results presented here do not depend on  
knowing the mass-temperature relation 
and its scatter, and we obtain a lower value of $\sigma_8 = 0.67^{+0.04}_{-0.05}$ for $\Omega_m = 0.3$. 
As shown in Figure~\ref{fig:bananaSpergel} our joint constraints on $\sigma_8$ and $\Omega_m$
agree very well with the 3-year WMAP data (Spergel et al.\ 2006). 
The 3-year WMAP data give a preference 
for $\sigma_8$ and $\Omega_m$ values lower than indicated by the first year WMAP data, combined with large scale structure data from SDSS 
(Tegmark et al.\ 2004). 
Our result also agrees well with many estimates of cluster abundances based on X-ray data 
(e.g., Reiprich \& B{\"o}hringer 2002; Seljak 2002; Viana, Nichol, \& Liddle 2002; Allen et al.\ 2003; Henry 2004), which have tended to produce $\sigma_8$ values in the lower end of the range 
of published estimates. Reiprich (2006) discusses cluster abundances in light of the 
3-year WMAP results, and gives best-fit results for $\sigma_8$ and $\Omega_m$ based on the 
HIFLUGCS cluster sample that agree very well with both WMAP data and the results presented here.
These results imply a lower value of $\sigma_8$ than implied by 
``cosmic shear'' measurements of weak gravitational lensing in random fields 
(Hoekstra et al.\ 2005; Semboloni et al.\ 2006; see figure~\ref{fig:bananaSpergel}).

One possible source of systematic error in the results presented in this paper is the use of X-ray luminosity selected clusters to estimate the 
scatter about the mass-X-ray luminosity relation. 
Some weak lensing studies (e.g., Erben et al.\ 2001; Dahle et al.\ 2003) may hint 
toward the existence of a class of X-ray under-luminous clusters, although detections of shear-selected clusters are 
significantly affected by projection effects (White, van Waerbeke, \& Mackey 2002; 
Hennawi \& Spergel 2005), 
and may include ``proto-clusters'' (Weinberg \& Kamionkowski 2002) that are still undergoing collapse and 
have not yet become fully virialised systems. The latter type of objects may pose less of a problem for the 
present study, since such systems should not be included when making the fit to a theoretically 
predicted mass function for fully collapsed systems. 
A sub-population of X-ray under-luminous clusters is also found by Popesso et al.\ (2006), based on a correlation
of Abell clusters in the SDSS database and RASS sources. Since these systems show signs of being in formation, still 
undergoing significant mass accretion, their relation to theoretically predicted abundances of virialised clusters
is somewhat unclear.   
Weak lensing surveys with the large sky coverage required to find statistically useful numbers of 
$M_{180c} \gtrsim 10^{15} M_{\sun}$ clusters 
are still some years into the future, but ongoing surveys (Miyazaki et al.\ 2003; Hetterscheidt et al.\ 2005; Gavazzi \& Soucail 2006; Wittman et al.\ 2006; Schirmer et al.\ 2006) are beginning to detect a significant number of lower-mass 
clusters based on weak gravitational shear. 
If there is indeed a large population of massive, X-ray under-luminous galaxy clusters, our 
estimates of sample incompleteness as a function of mass would be too low, implying a somewhat higher true value of $\sigma_8$. 

The constraints presented here may be improved in several ways in the near future: Firstly, obtaining new wide-field imaging data for all the 
clusters in this sample will remove the need for extrapolation of an adopted mass model out to $r_{180c}$, removing one potential source of systematic error while significantly reducing the 
statistical errors on the mass measurements. 
Furthermore, by obtaining imaging in a larger number of pass-bands, approximate photometric redshifts may be 
calculated, providing sufficient 3D information to distinguish the gravitationally lensed background galaxies from cluster and foreground galaxies. 
Although we have made an attempt to correct for this effect by applying an average, radially dependent, contamination correction
to all clusters (and estimate the extra scatter caused by cluster richness variations), 
the true correction factor will of course depend on cluster richness and cluster redshift, and cluster substructure will cause it not
to be a simple function of cluster radius. 
Hence, the appropriate correction for individual clusters can differ significantly from the 
average correction calculated here 
(see e.g., Broadhurst et al.\ 2005). Future photometric redshift estimates of background galaxies, and the uncertainties 
in these estimates could also be included in a optimized weighting scheme (e.g., Kaiser 2000; Benitez 2000), significantly improving the signal-to-noise ratio of the lensing mass measurements.   

The sample used in this study is restricted to galaxy clusters in the Northern celestial hemisphere, and the sample size could easily be doubled by 
including clusters with similar X-ray selection criteria in in the Southern hemisphere, e.g., from the REFLEX sample of B{\"o}hringer et al.\ (2004), 
thus improving the statistical errors. The sample could also be extended by lowering the X-ray luminosity cutoff $L_{\rm min}$, sampling 
the mass function at lower masses. For the (e)BCS sample, such a low-luminosity extension of the cluster sample would be restricted 
to the near end of the redshift range probed here, since $L_{\rm min}$ corresponds to the lower eBCS X-ray flux limit at the far end of the survey volume.    
However, more recent cluster surveys based on RASS data (e.g., Ebeling, Edge, \& Henry 2001) go down to 
lower X-ray fluxes and would be useful for probing further down the mass function without restricting the 
survey volume.   
Finally, by combining the data presented in this paper with a weak lensing survey of a similar cluster sample at higher redshifts, interesting constraints can also be put on the dark energy equation of state 
parameter $w$, based on the evolution of the cluster mass function. 

\section{Acknowledgments}

The author acknowledges support from the Research Council of Norway, including a postdoctoral research fellowship.
The author wishes to thank Doug Clowe, Harald Ebeling, {\O}ystein Elgar{\o}y, Pat Henry, Nick Kaiser, Per B.\ Lilje, Gerry Luppino and Gillian Wilson for helpful discussions related to this work and thanks Martin White for useful comments on a draft version of this manuscript. 
Matthew Bershady is thanked for providing the BCES regression software. 
The staffs of the University of Hawaii 2.24m telescope and the Nordic Optical Telescope are thanked for support 
during observing runs.  
Some of the data presented here have been taken using ALFOSC, which is owned by the 
Instituto de Astrofisica de Andalucia (IAA) and operated at the Nordic Optical Telescope 
under agreement between IAA and the NBIfAFG (Astronomical Observatory) of the University of Copenhagen.  
This research has made use of the NASA/IPAC Extragalactic Database (NED) which
is operated by the Jet Propulsion Laboratory, California Institute of 
Technology, under contract with the National Aeronautics and Space Administration.

{
}

\end{document}